\documentstyle[emulateapj]{article}

\begin{document}

\def\SN{signal-to-noise ratio}
\def\mass{$M_{\odot}$}
\def\lya{Ly$\alpha$}
\def\degree{$^{\circ}$}
\def\kps{km s$^{-1}$}
\newcommand{\eex}[1]{\times 10^{#1}}

\title{Constraints on Cool Gas in Rich Clusters of Galaxies}
\author{Eric D.\ Miller}
\affil{Dept.\ of Astronomy, University of Michigan,
Ann Arbor, MI
emiller@astro.lsa.umich.edu}
\author{Joel N.\ Bregman}
\affil{Dept.\ of Astronomy, University of Michigan,
Ann Arbor, MI
jbregman@umich.edu}
\author{Patricia M.\ Knezek}
\affil{STScI, Baltimore, MD
knezek@stsci.edu}

\begin{abstract}

Cool gas should be present in galaxy clusters due to 
stripping of galactic gas, infall onto the cluster, and from 
cooling flows.  We searched for this gas through resonance 
absorption line observations from low-ionization metal line 
gas toward six background quasars.  Both cooling flow and 
non-cooling flow clusters were observed, with lines of sight 
from the inner to outer parts of the cluster ($0.2 \leq 
r_{\rm proj}/R_{\rm cluster} \leq 0.7$). The HST/FOS spectra 
failed to detect Fe II or Mg II absorption at the cluster 
redshift, with 1-$\sigma$ upper limits on the ion column 
densities of $N \leq 10^{12}$--$10^{13}$\ cm$^{-2}$.  The 
failure to detect absorption lines implies that the cool 
cluster gas either is more highly ionized or it has a low 
covering factor.

\end{abstract}

\section{Introduction}

Rich clusters of galaxies, such as the Abell clusters, are filled with
$\sim 10^{14}$ \mass\ of hot ($T \sim 10^7$) X-ray emitting gas (see,
e.g., Fabian, Nulsen, and Canizares 1991).  The density of this gas is
largest in the central region (0.1 - 0.001 cm$^{-3}$) and decreases with
radius, where it becomes undetectable to X-ray instruments when n$_{\rm e}
< 10^{-4}$ cm$^{-3}$ (typically at radii of 1 - 3 Mpc; H$_{\rm o} = 50$
{\kps} Mpc$^{-1}$ throughout).  Although studies of cluster gas have focused
on the hot medium, three important processes can 
produce cool gas ($T \sim 10^3$--$10^4$ K) in rich clusters of galaxies: stripping of galactic gas by
the hot ambient medium; infall of cool gas into the cluster; and the
radiative cooling of the hot gas in the cluster core (cooling flows).

The stripping of galactic gas is expected to be the main process by which
the cluster gas becomes enriched, so it is central to interpreting cluster
metallicities.  The rate at which gas is removed from the ensemble of 
galaxies is $\sim 100$ \mass\ yr$^{-1}$ Soker, Bregman, and Sarazin 1991), this gas should be near solar metallicity, and the stripping should occur in the inner 1 Mpc of the 
cluster, where the ambient medium is densest (Gaetz, Salpeter, and Shaviv 1987).  The infall of gas onto
clusters occurs at the outer regions (3 Mpc radius), it will involve low 
metallicity material, and the infall rate is expected to be 
$\sim 10^3 - 10^4$ \mass\ yr$^{-1}$.  The growth rate of the cluster by 
infalling gas is a prime growth mechanism for clusters.  Finally, in the
central 100 kpc, clusters with cooling flows are believed to deposit cooled 
gas of approximately 1/3 solar metallicity and at the rate of 
$\sim 100$ \mass\ yr$^{-1}$.  Each of these mechanisms is distinct both
in the spatial distribution for the absorption and in the metallicity of 
the gas.  This paper is the first in an effort to study this cooled 
gas through the presence of absorption lines.

There have been studies at a variety of wavebands in an effort to detect
cool gas.  Direct detection of HI in absorption and emission have met
with some success, although large quantities of gas are not discovered,
and CO emission remains undetected 
(McNamara, Bregman, and O'Connell 1990; McNamara and Jaffe
1994; O'Dea, Baum, and Gallimore 1994; O'Dea et al. 1994; O'Dea, Gallimore,
and Baum 1995; O'Dea, Payne, and Kocevski 1998).  
Indirect evidence for cool gas has been obtained by far infrared studies
that are sensitive to thermal reemission from dust, where weak emission is
seen in a minority of galaxy clusters 
(Maoz 1995; Cox, Bregman, and Schombert 1995; Stickel et al. 1998). 
Large amounts of cooled gas ($10^{12}$ \mass) are inferred from the excess
soft X-ray absorption toward the central regions (within $r < 0.5$ Mpc) of
cooling flow clusters (White et al.\ 1991; Allen et al.\ 1993; Fabian et al.\
1994; Fukazawa et al.\ 1995), although this result has been questioned
(Arabadjis and Bregman 1999).

Cooled material is detectable through UV and optical
absorption line studies against background point sources.  
In a recent study, Koekemoer et al.\ (1998) searched for absorption against 
a quasar associated with the central galaxy of Abell 1030.  No lines were
detected and they obtain 1-$\sigma$\ upper limits on
the column density for a number of species, including HI ($N < 10^{12.4}$
cm$^{-2}$), MgII ($N < 10^{11.5}$ cm$^{-2}$), and FeII ($N < 10^{12.0}$
cm$^{-2}$).  In another study,
toward NGC 1275 in Perseus (Johnstone and Fabian 1995), there is a dip in the middle of the \lya\ emission line at the velocity of the 21 cm HI absorption feature, which may be interpreted as \lya\ absorption with an equivalent width in the 1-5 \AA range.  However, Johnstone and Fabian (1995) interpret the \lya\ emission line as being double but without absorption, which illustrates the difficulty of analyzing emission and absorption features at the same redshift.

The problem with using an AGN in the center of a cluster is that if 
absorption is detected, it is difficult to distinguish absorption in
the AGN from cluster absorption.  Alternatively, a failure to detect 
absorption might be due to the strong photoionizing influence of the
AGN on the cool gas.
Both problems are avoided by detecting
cool ICM gas through absorption against background AGNs which are
unrelated to the cluster.  This technique has the desirable feature that
the cluster and background source are well separated in velocity space,
allowing clean identification of spectral features from each.  In addition,
the path length covers the entire cluster, thereby maximizing the
likelihood of detection.  We have obtained near-UV spectra of five 
background QSOs behind different clusters, supplemented by an additional 
QSO spectrum from the HST archive.  With these six independent lines 
of sight at various distances from the cluster cores, we have searched for
low ionization metal resonance lines.

\section{Target Selection and Observations}

The background AGNs were identified by us as part of a program to find
continuum point sources positioned behind galaxies and clusters of
galaxies (Knezek and Bregman 1998).  Our procedure was to use archived
ROSAT X-ray images of nearly sixty northern hemisphere clusters and
identify point sources, many of which are background AGNs.  We obtained
optical spectrophotometry of the sources with the MDM 2.4m telescope, and
when AGNs were confirmed through their emission line spectra, we determined
their redshifts, fluxes and spectral shapes.  

The confirmed AGNs were screened by brightness, redshift, reddening, and
projected distance from the cluster center, and the best targets were
chosen for observation with HST.  In particular, to allow for unambiguous
detection of the MgII $\lambda 2800$ doublet, we limited the redshift to $z
< 1.3$.  At redshifts greater than this, the Ly$\alpha$\ forest is shifted
past the MgII doublet, which could cause false detections (this is a 
problem for the sixth target, taken from the archive, as it has $z > 1.3$).

The final sample is unbiased and representative of clusters as a whole, as our selection
of AGNs samples a range of projected radii and X-ray properties.  Two of
the AGNs lie within the inner one-fifth of the cluster radius, while the
other four lie between 0.3--0.72 $R_{\rm cluster}$.  With regards to the
X-ray properties, one of the clusters, Abell 1795, exhibits a very large
cooling rate while Abell 754 has a more modest flow.  Two of the clusters,
Abell 21 and Coma, show no evidence of cooling flows.  The remaining two,
Abell 151 and Abell 1267, do not have public X-ray data, although they are
nearby and should be bright.  Finally, our sample is unbiased by the
selection of AGNs, since these background sources are completely
uncorrelated to the clusters themselves.

The data for this project were obtained with the Faint
Object Spectrograph (FOS).  Each spectrum was taken through the 0.86''
circular aperture using the G270H grating and Red Digicon detector.  In
this configuration, the spectrum covers the 2222--3277 \AA\ range, and the
expected FOS instrumental line width (FWHM) is 2.04 \AA, which produces a
spectral resolution of 219 \kps\ at 2800 \AA.

The observations are summarized in Table 1, which lists the date and
exposure time for each observation, as well as characteristics of each QSO
and cluster.  Five of the observations were obtained for this program.  The
sixth, QSO 1258$+$285 behind the Coma cluster, was obtained from the HST Data
Archives to increase our sample.  Additional GHRS spectra in the far-UV
were also retrieved, but were found to be of poor quality for this project.

Each data set was recalibrated with the CALFOS routine in the IRAF/STSDAS
package, using the best reference files available.  This produced a
substantial improvement (about a factor of 1.5) in the signal-to-noise over
the initial pipeline processing.  The calibrated spectra
along with the propagated photon counting errors are shown in Figure 1; important QSO emission lines are marked.

The identification of absorption lines and measurement of their strengths
were performed using the HST Absorption Line Key Project software
(Schneider et al.\ 1993).  First, a QSO ``continuum'' was fit to each
spectrum with a series of spline curves.  This fit included the QSO
emission lines to allow detection of absorption lines superimposed on these
features; each spectrum and associated error array was normalized to
unity.  The line-searching software takes as input parameters the maximum
Gaussian width (FWHM), detection limit (defined in terms of the
significance level of the measured equivalent width), and the PSF
characteristics.  The software performs both PSF and Gaussian fitting to
the input spectrum, applying multiple Gaussians in the case of blended
features.  (A detailed description of the line-searching algorithm is
provided by Schneider et al.\ 1993.) We constrained the FWHM of an
absorption line to lie between the instrumental limit and that
corresponding to a velocity dispersion of 1000 \kps, which is greater
than the velocity dispersion of these clusters.  A detection threshold 
of 3-$\sigma$ was used.

Several lines were detected in each spectrum, and these are
listed in Table 2 along with their measured equivalent widths and our
identifications.
The strong Galactic lines of FeII ($\lambda 2344$, $\lambda 2374$, $\lambda
2382$, $\lambda 2586$, and $\lambda 2600$), MgII ($\lambda\lambda
2796,2803$), and MgI ($\lambda 2853$) appear in every spectrum, with line
strengths ranging from 0.2 \AA\ for the MgI line to about 1.0 \AA\ for the
stronger line of the MgII doublet (Figure 1; all wavelengths are in
vacuum.) Analysis of the Galactic
features sheds some light on the quality of the wavelength calibration.
The wavelength zeropoint of FOS was not constant between observations, 
but varied as a result of non-repeatability of the filter/grating wheel
position, in addition to other effects.  Since contemporaneous wavelength
calibration data were not taken with any of the object spectra, the
wavelength zeropoint could be be displaced by as much as 250 \kps (Keyes, HST
Handbook).  Indeed, the line centers of the Galactic lines are shifted from
rest by as much as $\pm 2.0$ \AA\ ($\pm 225$ \kps) in our spectra. Since it
is impossible to separate true Galactic motion from errors in the
wavelength zeropoint, we have adopted the convention of the HST key project
team and have shifted each spectrum so that the Milky Way lines are at
rest, using the strong FeII and MgII lines as reference.  All figures and
data reported in this paper reflect this adjustment.

In addition to the Galactic lines, several other features have been
identified in the spectra, and are described below.  No sets of lines
within $\pm 3000$ \kps\ of the cluster redshift were identified for any of
the clusters.  The spectra in Figure 1 show the regions in which we would
expect to see the MgII or FeII absorption produced by cool cluster gas.

To detect fainter absorption features, we searched each spectrum using a
detection threshold of 1.5-$\sigma$.  Detected lines falling within the
expected regions were compared in velocity and line strength to see if they
were plausible components of the same species.  No satisfactory matches
were made.

\section{Analysis}

\subsection{Absorption Systems}

We see absorption by several systems which are unrelated to the intervening
Abell clusters or to the Galaxy.  

{\it QSO 0107$-$156\/}: We observe a $W_{\lambda} \approx 1.0$ \AA\
absorption line at 2244.8 \AA, superimposed on the QSO Ly$\alpha$\ emission
line.  This could be due to an HI cloud in the AGN itself, in
which case the cloud would be outflowing at $-$1630 \kps\ relative to the
QSO.

This spectrum contains two additional absorption lines that are more difficult
to identify:  $W_{\lambda} = 0.94$\ \AA\ at 2397.5 \AA\ and
$W_{\lambda} = 0.63$\ \AA\ at 2402.0 \AA.  In the rest frame of the
cluster, these lines would fall at 2277.7 and 2282.0 \AA, respectively; in
the rest frame of the QSO, they would be 1739.8 and 1743.1 \AA.  There are
no known resonance lines at these wavelengths.  If this absorption were due
to intervening material, e.g. from FeII $\lambda 2374$ or $\lambda 2382$,
we would expect to see absorption elsewhere in the spectrum from species in
a similar ionization state.  (In the example, FeII $\lambda 2600$ would
fall in the range 2615--2630 \AA\ and have $W_{\lambda} > 0.7$ \AA; the
1-$\sigma$ equivalent width limit, found below, is 0.06 \AA\ in this region
of the spectrum.)

The most likely identification for this pair of lines is the C IV doublet
at 1548 \AA\ and 1550 \AA\ due to an intervening redshift system at z = 0.549.
The separation between the pair of absorption lines is 4.5 \AA, while the 
redshifted separation of the C IV doublet is 4.00 \AA, a 3.6 $\sigma$
difference.  Also, the ratio of the two lines, $W_{2402}/W_{2397} = 0.67$,
is consistent with the ratio from the C IV doublet in the optically thin
limit.  Unfortunately, we cannot confirm this suggestion by observing other
high-ionization lines, since the strong lines lie below the wavelength
limit of our observation.   The only other line pair with the same fraction
separation as the observed absorption ines are the OI $\lambda 1302$ and the
SiII $\lambda 1304$ lines, which would have a redshift of 0.841.  However,
we rule out this identification due to the absence of other strong
low-ionization absorption lines at the same redshift.

{\it QSO 1127$+$269\/}:  We discovered strong FeII and MgII absorption at
$z = 0.152$ from a system between this QSO and the foreground cluster.
These features are noted by tick marks in Figure 1.  Imaging at V and I bands by us with the MDM 1.3m telescope
yielded no evidence of faint structure superimposed on or near the QSO.

{\it QSO 1258$+$285\/}: This spectrum appears to contain several Ly forest
lines blueward of the QSO Ly$\alpha$\ and Ly$\beta$\ emission lines.  It is
difficult to identify absorption from the intervening cluster gas against
this forest.  In particular, the region where we would expect to find MgII
absorption lies directly on the Ly$\alpha$\ emission from the QSO, making
it impossible to separate true absorption from structure in the emission
line.  There appears to be absorption in some of the regions where we
expect to find some of the FeII absorption lines within the cluster.
However, one of the strongest Fe lines, FeII
$\lambda 2600$\ is undetected, so we conclude that Fe II absorption at the
cluster redshift is unlikely.  Absorption at the location of several 
other Fe II lines are probably Ly forest lines.

Since the QSO Ly$\beta$\ line is shifted into our wavelength region, we can
attempt to match corresponding Ly$\beta$\ and Ly$\alpha$\ forest lines.
For example, the doublet seen at 2280 \AA\ is repeated 2700 \AA, making
this a likely Ly$\beta$/Ly$\alpha$\ pair.  Other such pairs are observed at
2320 \AA/2750 \AA and 2415 \AA/2860 \AA, the latter of these being strong
absorption near or within the QSO.  The features seen between 2450--2620
\AA\ could be Ly$\alpha$\ forest lines, however their Ly$\beta$
counterparts would lie below our short-wavelength cutoff.  For redshift
determination of these and other ambiguous features in Table 2, we assume
they are Ly$\alpha$\ lines.

\subsection{ICM Metal Absorption Limits}

We can place an upper limit on the equivalent width of any expected
absorption lines simply by using the noise calculations performed by the
line-searching software.  The upper limits correspond to a maximum
velocity dispersion of $\sigma = 1000$ \kps and are such that these lines,
if present, would fall on the linear portion of the curve of growth.  The
column density for a weak (optically thin) line can be determined in the
usual fashion using the relation (see, e.g., Spitzer 1978)

\[ N = 1.13 \eex{20} W_{\lambda} ({\rm \AA}) /
   f \lambda^{2} ({\rm \AA})~~{\rm cm}^{-2}\]

After correcting the equivalent width limits for redshift, we found upper
limits to the column density of each species with lines falling in our
spectral range.  We employed a weighted average over all lines of a given
species to further constrain the column density, weighting by the expected
line strength and then adding individual line contributions in quadrature.
The resulting 1-sigma upper limits to the column densities for each cluster
are listed in Table 3, along with the wavelength and $f$-value of the
strongest expected transition of each species.

\section{Conclusions}

We have placed strict constraints on the amount of low-ionization absorbing
material present in rich clusters of galaxies.  If we assume a large
covering fraction, then the column densities of the species in Table 2 are
$N \leq 10^{12}$ -- $10^{13}$ cm$^{-2}$, and there is little low-ionization
cool gas in the ICM.  Our results are consistent with those of Koekemoer et
al. (1998), who found similar limits to $N$ for a variety of species along
a single line of sight in Abell 1030.

The results permit us to place constraints on the covering fraction of cool
absorbing gas. With six independent lines of sight producing non-detections
for the strongest expected line (MgII $\lambda2796$), the covering fraction
of low-ionization gas with $N({\rm MgII}) \geq 3\eex{12}$ cm$^{-2}$ must be
less than 40\%, at the 95\% confidence level.

Although low ionization gas does not appear to have a large covering factor
in galaxy clusters, little is known about higher ionization state gas,
as is often identified through CIV and SiIV absorption.  There are
expectations that searches with the higher ionization will be more successful,
based upon a few observations through other clusters.
In an observation toward 3C~273, which lies behind the outer part of the
Virgo cluster, Ly$\alpha$ absorption was detected at the redshift of the
Virgo cluster (Bahcall et al.\ 1991).  
It has a HI column density of at least 4$\times$10$^{14}$
cm$^{-2}$ and has detectable high ionization metal absorption lines
(Hurwitz et al. 1998 and references therein).
In the future, we hope to study whether clusters are sites of high--
ionization absorption lines, and whether they produce Ly$\alpha$ absorption,
thus completing the study of the absorbing properties of galaxy clusters.

We would like to thank Don Schneider, Jane Charleton, and Kaspar von Braun 
for their insights and assistance.  Financial support was provided by 
NASA through the HST GO program.

\clearpage

\begin{figure}
\centerline{\epsfbox{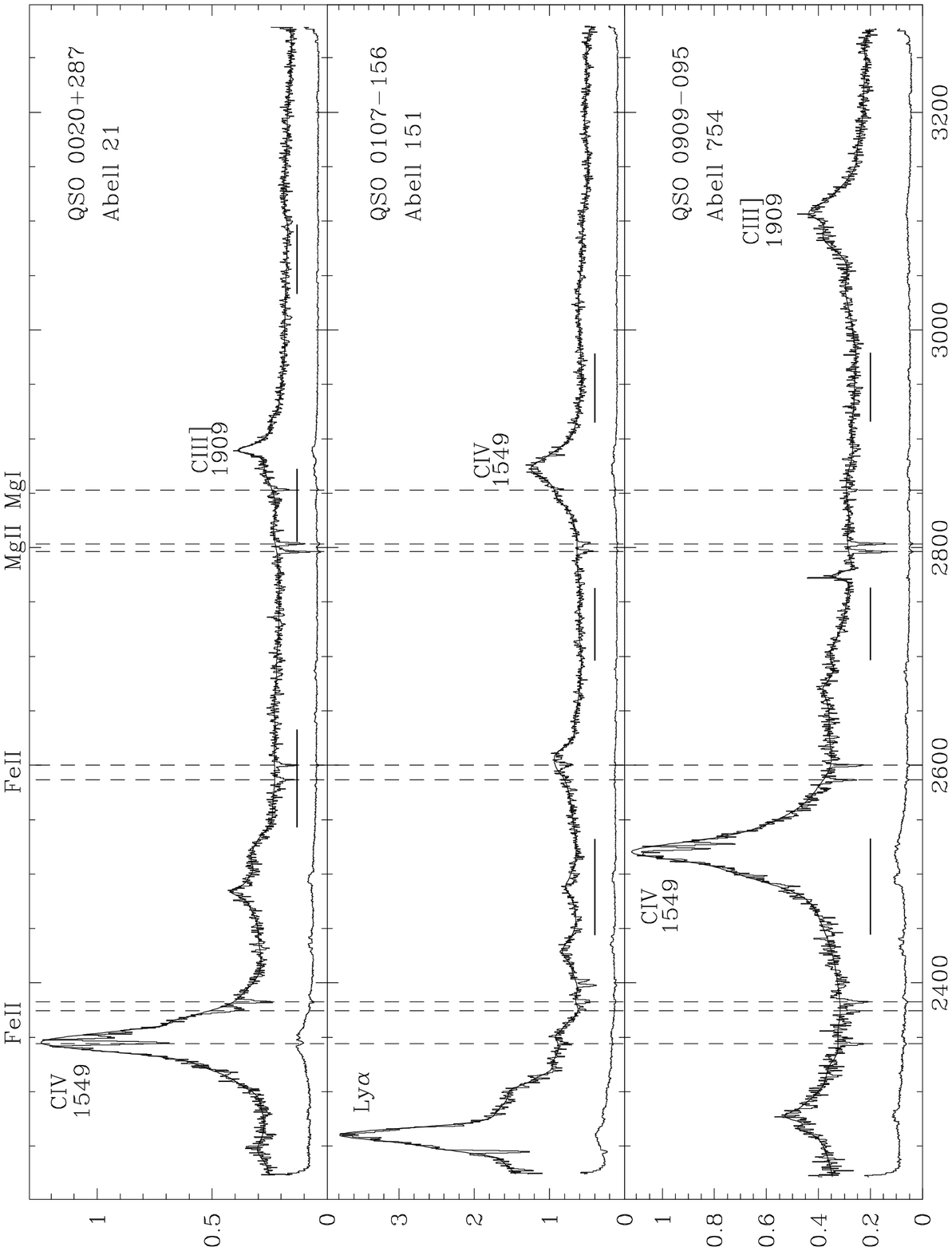}}
\label{fig1}
\end{figure}

\clearpage

\begin{figure}
\centerline{\epsfbox{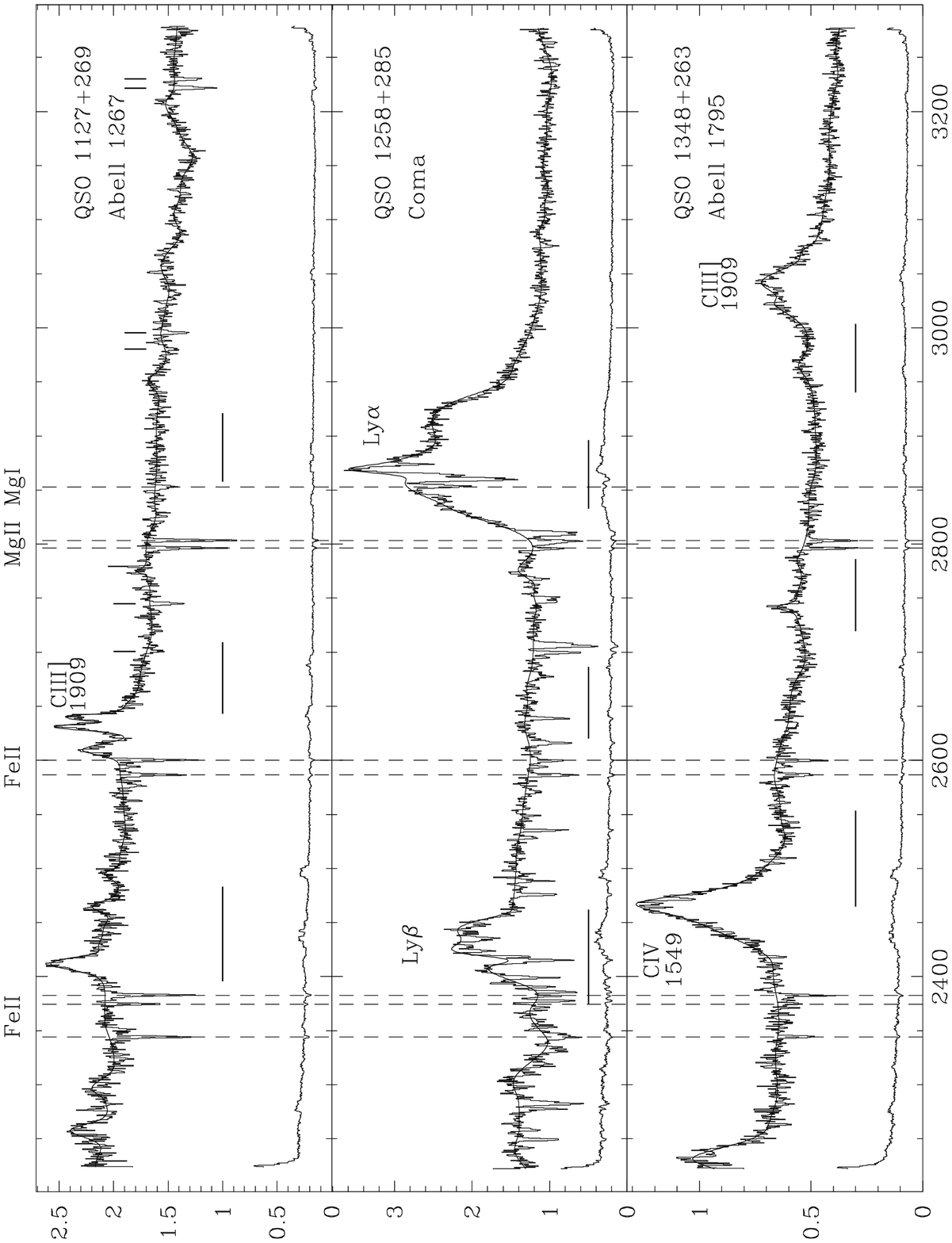}}
\label{fig2}
\end{figure}

\clearpage

\begin{tabular}{|l|l|c|c||l|l|c|c|c|}
\multicolumn{9}{c}{\sc Table 1} \\
\hline \hline
QSO & 
$z_{\rm qso}$ & 
$t_{\rm exp}$ &
Date Obs. &
Cluster & 
$z_{\rm cluster}$ & 
$d_{\rm proj} \over R_{\rm cluster}$ & 
$R_{\rm cluster}$  & 
${\dot M}$ \\ 
 & 
 & 
(s) &
 & 
 &
 &
 &
(Mpc) & 
${\rm M_{\odot} \over yr}$ \\
\hline
0020$+$287 & 0.51  & 7820 & 17 Nov 1996 & Abell 21   & 0.0948 & 0.19 & 2.65 & 0 \\
0107$-$156 & 0.861 & 3970 & 27 Jul 1996 & Abell 151  & 0.0526 & 0.40 & 2.30 & - \\
0909$-$095 & 0.63  & 8740 & 11 Feb 1997 & Abell 754  & 0.0528 & 0.31 & 2.58 & 24 \\
1127$+$269 & 0.378 & 4050 & 28 Nov 1996 & Abell 1267 & 0.0321 & 0.19 & 1.68 & - \\
1258$+$285 & 1.36  & 2100 &  5 Apr 1995 & Abell 1656 & 0.0232 & 0.35 & 2.30 & 0 \\
1348$+$263 & 0.59  & 6410 & 22 Jan 1997 & Abell 1795 & 0.0616 & 0.72 & 1.94 & 295 \\
\hline
\end{tabular}

\begin{tabular}{|c|c|c|l|r|}
\multicolumn{5}{c}{\sc Table 2} \\
\hline \hline
$\lambda_{\rm obs}$ (\AA) & $W_{\lambda}$ (\AA) & FWHM (\AA) & 
\multicolumn{1}{c|}{ID} & $v$ or $z$ \\
\hline \hline
\multicolumn{5}{|l|}{QSO 0020$+$287} \\
\hline
2345.00 $\pm$ 0.06 & 1.51 $\pm$ 0.06 & 3.09 $\pm$ 0.15 & 
 CIV 1548.19 (QSO) & 0.515 \\
2350.01 $\pm$ 0.14 & 0.48 $\pm$ 0.07 & 2.16 $\pm$ 0.36 & 
 CIV 1550.77 (QSO) & 0.515 \\
2353.78 $\pm$ 0.31 & 0.34 $\pm$ 0.08 & 2.70 $\pm$ 0.81 & 
 ... & ... \\
2375.10 $\pm$ 0.22 & 1.18 $\pm$ 0.12 & 4.50 $\pm$ 0.54 & 
 FeII 2374.46 (MW) &     80.29 \\
2382.88 $\pm$ 0.09 & 0.95 $\pm$ 0.08 & 2.05 $\pm$ 0.21 & 
 FeII 2382.76 (MW) &     14.01 \\
2578.39 $\pm$ 0.40 & 0.50 $\pm$ 0.13 & 3.10 $\pm$ 0.95 & 
 ... & ... \\
2586.48 $\pm$ 0.27 & 1.10 $\pm$ 0.15 & 4.05 $\pm$ 0.63 & 
 FeII 2586.65 (MW) &    -20.00 \\
2600.13 $\pm$ 0.11 & 0.95 $\pm$ 0.10 & 2.19 $\pm$ 0.27 & 
 FeII 2600.17 (MW) &     -4.61 \\
2796.28 $\pm$ 0.06 & 1.41 $\pm$ 0.09 & 2.07 $\pm$ 0.15 & 
 MgII 2796.35 (MW) &     -8.20 \\
2803.57 $\pm$ 0.07 & 1.40 $\pm$ 0.09 & 2.21 $\pm$ 0.17 & 
 MgII 2803.53 (MW) &      4.58 \\
2815.76 $\pm$ 0.35 & 0.27 $\pm$ 0.08 & 2.04 $\pm$ 0.00 & 
 ... & ... \\
2853.49 $\pm$ 0.15 & 0.61 $\pm$ 0.07 & 2.04 $\pm$ 0.00 & 
 MgI 2852.96 (MW) &     55.01 \\
3011.57 $\pm$ 0.30 & 0.46 $\pm$ 0.12 & 2.43 $\pm$ 0.72 & 
 ... & ... \\
\hline
\multicolumn{5}{|l|}{QSO 0107$-$156} \\
\hline
2244.79 $\pm$ 0.07 & 0.98 $\pm$ 0.06 & 2.14 $\pm$ 0.16 & 
  HI 1215.67 (QSO) & 0.847 \\
2257.28 $\pm$ 0.28 & 0.19 $\pm$ 0.04 & 2.02 $\pm$ 0.00 & 
 ... & ... \\
2263.05 $\pm$ 0.47 & 0.32 $\pm$ 0.08 & 3.99 $\pm$ 1.14 & 
 ... & ... \\
2268.67 $\pm$ 0.26 & 0.25 $\pm$ 0.05 & 2.02 $\pm$ 0.00 & 
 ... & ... \\
2344.14 $\pm$ 0.42 & 0.53 $\pm$ 0.13 & 3.62 $\pm$ 1.02 & 
 FeII 2344.21 (MW) &    -10.04 \\
2382.83 $\pm$ 0.17 & 0.95 $\pm$ 0.12 & 2.83 $\pm$ 0.42 & 
 FeII 2382.76 (MW) &      7.75 \\
2397.53 $\pm$ 0.11 & 0.94 $\pm$ 0.10 & 2.16 $\pm$ 0.27 & 
 CIV 1548.20 & 0.549 \\
2402.00 $\pm$ 0.15 & 0.62 $\pm$ 0.08 & 2.02 $\pm$ 0.00 & 
 CIV 1550.77 & 0.549 \\
2577.41 $\pm$ 0.34 & 0.24 $\pm$ 0.07 & 2.04 $\pm$ 0.00 & 
 MnII 2576.88 (MW) &     62.07 \\
2600.01 $\pm$ 0.22 & 1.14 $\pm$ 0.11 & 4.63 $\pm$ 0.55 & 
 FeII 2600.17 (MW) &    -18.66 \\
2607.00 $\pm$ 0.43 & 0.36 $\pm$ 0.10 & 3.29 $\pm$ 1.04 & 
 MnII 2606.46 (MW) &     61.94 \\
2778.91 $\pm$ 0.37 & 0.23 $\pm$ 0.07 & 2.04 $\pm$ 0.00 & 
 ... & ... \\
2796.35 $\pm$ 0.14 & 0.95 $\pm$ 0.10 & 2.88 $\pm$ 0.35 & 
 MgII 2796.35 (MW) &     -0.25 \\
2803.60 $\pm$ 0.20 & 0.89 $\pm$ 0.11 & 3.43 $\pm$ 0.48 & 
 MgII 2803.53 (MW) &      7.61 \\
2884.86 $\pm$ 0.29 & 0.26 $\pm$ 0.06 & 2.04 $\pm$ 0.00 & 
 ... & ... \\
3190.84 $\pm$ 0.36 & 0.25 $\pm$ 0.07 & 2.04 $\pm$ 0.00 & 
 ... & ... \\
\hline
\multicolumn{5}{|l|}{QSO 0909$-$095} \\
\hline
2286.25 $\pm$ 0.37 & 0.39 $\pm$ 0.11 & 2.60 $\pm$ 0.89 & 
 ... & ... \\
2314.85 $\pm$ 0.37 & 0.30 $\pm$ 0.09 & 2.02 $\pm$ 0.00 & 
 ... & ... \\
2344.14 $\pm$ 0.15 & 0.77 $\pm$ 0.11 & 2.17 $\pm$ 0.36 & 
 FeII 2344.21 (MW) &     -9.23 \\
2374.19 $\pm$ 0.30 & 0.72 $\pm$ 0.14 & 3.29 $\pm$ 0.72 & 
 FeII 2374.46 (MW) &    -34.20 \\
2382.72 $\pm$ 0.17 & 1.01 $\pm$ 0.12 & 2.94 $\pm$ 0.41 & 
 FeII 2382.76 (MW) &     -5.05 \\
2489.35 $\pm$ 0.32 & 0.31 $\pm$ 0.09 & 2.38 $\pm$ 0.77 & 
 ... & ... \\
2523.34 $\pm$ 0.09 & 0.67 $\pm$ 0.05 & 2.44 $\pm$ 0.22 & 
 CIV 1548.19 (QSO) & 0.630 \\
2527.91 $\pm$ 0.09 & 0.56 $\pm$ 0.04 & 2.02 $\pm$ 0.00 & 
 CIV 1550.77 (QSO) & 0.630 \\
2586.57 $\pm$ 0.10 & 0.76 $\pm$ 0.06 & 2.02 $\pm$ 0.00 & 
 FeII 2586.65 (MW) &     -9.90 \\
2599.98 $\pm$ 0.10 & 0.80 $\pm$ 0.06 & 2.02 $\pm$ 0.00 & 
 FeII 2600.17 (MW) &    -22.62 \\
2796.46 $\pm$ 0.07 & 1.25 $\pm$ 0.08 & 2.25 $\pm$ 0.18 & 
 MgII 2796.35 (MW) &     11.97 \\
2803.62 $\pm$ 0.07 & 1.17 $\pm$ 0.08 & 2.17 $\pm$ 0.18 & 
 MgII 2803.53 (MW) &      9.44 \\
2852.93 $\pm$ 0.29 & 0.37 $\pm$ 0.09 & 2.34 $\pm$ 0.69 & 
 MgI 2852.96 (MW) &     -3.23 \\
3086.73 $\pm$ 0.34 & 0.21 $\pm$ 0.06 & 2.04 $\pm$ 0.00 & 
 ... & ... \\
\hline
\end{tabular}
\newpage
\begin{tabular}{|c|c|c|l|r|}
\multicolumn{5}{c}{\sc Table 2 (cont.)} \\
\hline \hline
$\lambda_{\rm obs}$ (\AA) & $W_{\lambda}$ (\AA) & FWHM (\AA) & 
\multicolumn{1}{c|}{ID} & $v$ or $z$ \\
\hline \hline
\multicolumn{5}{|l|}{QSO 1127$+$269} \\
\hline
2252.05 $\pm$ 0.62 & 0.35 $\pm$ 0.10 & 4.45 $\pm$ 1.49 & 
 ... & ... \\
2344.13 $\pm$ 0.07 & 0.85 $\pm$ 0.06 & 2.12 $\pm$ 0.17 & 
 FeII 2344.21 (MW) &    -11.16 \\
2374.52 $\pm$ 0.12 & 0.51 $\pm$ 0.04 & 2.04 $\pm$ 0.00 & 
 FeII 2374.46 (MW) &      7.22 \\
2382.90 $\pm$ 0.07 & 0.88 $\pm$ 0.06 & 2.17 $\pm$ 0.16 & 
 FeII 2382.76 (MW) &     16.49 \\
2576.67 $\pm$ 0.28 & 0.30 $\pm$ 0.06 & 2.77 $\pm$ 0.66 & 
 MnII 2576.88 (MW) &    -23.68 \\
2586.55 $\pm$ 0.08 & 0.76 $\pm$ 0.05 & 2.25 $\pm$ 0.18 & 
 FeII 2586.65 (MW) &    -11.09 \\
2594.18 $\pm$ 0.25 & 0.21 $\pm$ 0.04 & 2.04 $\pm$ 0.00 & 
 MnII 2594.50 (MW) &    -37.41 \\
2600.29 $\pm$ 0.05 & 0.88 $\pm$ 0.05 & 2.12 $\pm$ 0.13 & 
 FeII 2600.17 (MW) &     13.76 \\
2606.33 $\pm$ 0.29 & 0.16 $\pm$ 0.04 & 2.04 $\pm$ 0.00 & 
 MnII 2606.46 (MW) &    -14.71 \\
2629.14 $\pm$ 0.32 & 0.20 $\pm$ 0.05 & 2.63 $\pm$ 0.77 & 
 ... & ... \\
2700.85 $\pm$ 0.30 & 0.21 $\pm$ 0.06 & 2.35 $\pm$ 0.70 & 
 FeII 2344.21  &  0.152 \\
2744.96 $\pm$ 0.12 & 0.45 $\pm$ 0.05 & 2.12 $\pm$ 0.28 & 
 FeII 2382.76  &  0.152 \\
2796.32 $\pm$ 0.05 & 0.98 $\pm$ 0.04 & 2.04 $\pm$ 0.00 & 
 MgII 2796.35 (MW) &     -3.50 \\
2803.49 $\pm$ 0.05 & 1.06 $\pm$ 0.03 & 2.04 $\pm$ 0.00 & 
 MgII 2803.53 (MW) &     -4.36 \\
2852.72 $\pm$ 0.23 & 0.36 $\pm$ 0.06 & 2.78 $\pm$ 0.54 & 
 MgI 2852.96 (MW) &    -25.16 \\
2980.19 $\pm$ 0.32 & 0.20 $\pm$ 0.06 & 2.33 $\pm$ 0.77 & 
 FeII 2586.65  &  0.152 \\
2995.51 $\pm$ 0.15 & 0.39 $\pm$ 0.05 & 2.21 $\pm$ 0.35 & 
 FeII 2600.17  &  0.152 \\
3031.19 $\pm$ 0.62 & 0.29 $\pm$ 0.08 & 4.50 $\pm$ 1.48 & 
 ... & ... \\
3204.64 $\pm$ 0.36 & 0.15 $\pm$ 0.04 & 2.04 $\pm$ 0.00 & 
 ... & ... \\
3221.81 $\pm$ 0.09 & 0.63 $\pm$ 0.06 & 2.22 $\pm$ 0.22 & 
 MgII 2796.35  &  0.152 \\
3230.12 $\pm$ 0.13 & 0.41 $\pm$ 0.04 & 2.04 $\pm$ 0.00 & 
 MgII 2803.53  &  0.152 \\
3273.44 $\pm$ 0.61 & 0.39 $\pm$ 0.11 & 4.28 $\pm$ 1.44 & 
 ... & ... \\
\hline
\multicolumn{5}{|l|}{QSO 1258$+$285} \\
\hline
2249.92 $\pm$ 0.12 & 0.95 $\pm$ 0.11 & 2.19 $\pm$ 0.30 & 
 Ly$\beta$ & 1.194 \\
2258.94 $\pm$ 0.29 & 0.40 $\pm$ 0.09 & 2.04 $\pm$ 0.00 & 
 Ly$\beta$ & 1.203  \\
2278.16 $\pm$ 0.14 & 0.86 $\pm$ 0.09 & 2.02 $\pm$ 0.00 & 
 Ly$\beta$ & 1.222  \\
2283.26 $\pm$ 0.11 & 2.08 $\pm$ 0.14 & 3.46 $\pm$ 0.28 & 
 Ly$\beta$ & 1.227  \\
2310.31 $\pm$ 0.32 & 0.49 $\pm$ 0.12 & 2.64 $\pm$ 0.75 & 
 Ly$\alpha$ & 0.900 \\
2319.17 $\pm$ 0.37 & 1.07 $\pm$ 0.17 & 4.72 $\pm$ 0.91 & 
 Ly$\beta$ & 1.262  \\
2332.41 $\pm$ 0.34 & 1.04 $\pm$ 0.17 & 4.28 $\pm$ 0.82 & 
 Ly$\alpha$ & 0.918 \\
2344.99 $\pm$ 0.20 & 1.92 $\pm$ 0.17 & 4.73 $\pm$ 0.49 & 
 FeII 2344.21 (MW) &     98.66 \\
2354.03 $\pm$ 0.33 & 0.35 $\pm$ 0.09 & 2.02 $\pm$ 0.00 & 
 Ly$\alpha$ & 0.936 \\
2372.10 $\pm$ 0.36 & 0.31 $\pm$ 0.09 & 2.02 $\pm$ 0.00 & 
 FeII 2374.46 (MW) &   -298.58 \\
2378.09 $\pm$ 0.12 & 0.84 $\pm$ 0.07 & 2.02 $\pm$ 0.00 & 
 Ly$\alpha$ & 0.956 \\
2382.73 $\pm$ 0.23 & 0.47 $\pm$ 0.08 & 2.02 $\pm$ 0.00 & 
 FeII 2382.76 (MW) &     -4.74 \\
2386.39 $\pm$ 0.10 & 0.97 $\pm$ 0.07 & 2.02 $\pm$ 0.00 & 
 Ly$\alpha$ & 0.963 \\
2391.17 $\pm$ 0.21 & 0.47 $\pm$ 0.08 & 2.02 $\pm$ 0.00 & 
 Ly$\alpha$ & 0.967 \\
2399.68 $\pm$ 0.08 & 1.05 $\pm$ 0.08 & 2.17 $\pm$ 0.20 & 
 Ly$\beta$ & 1.340  \\
2415.73 $\pm$ 0.07 & 1.36 $\pm$ 0.09 & 2.32 $\pm$ 0.18 & 
 Ly$\beta$ & 1.356  \\
2419.14 $\pm$ 0.13 & 0.64 $\pm$ 0.06 & 2.02 $\pm$ 0.00 & 
 Ly$\beta$ & 1.359  \\
2424.40 $\pm$ 0.16 & 0.54 $\pm$ 0.10 & 2.04 $\pm$ 0.38 & 
 Ly$\beta$ & 1.364  \\
2435.65 $\pm$ 0.27 & 0.88 $\pm$ 0.13 & 3.94 $\pm$ 0.68 & 
 Ly$\alpha$ & 1.003 \\
2447.94 $\pm$ 0.38 & 0.46 $\pm$ 0.10 & 3.39 $\pm$ 0.90 & 
 Ly$\alpha$ & 1.013 \\
2476.42 $\pm$ 0.09 & 0.98 $\pm$ 0.09 & 2.26 $\pm$ 0.23 & 
 Ly$\alpha$ & 1.037 \\
2488.99 $\pm$ 0.14 & 0.64 $\pm$ 0.07 & 2.04 $\pm$ 0.00 & 
 Ly$\alpha$ & 1.047 \\
2529.08 $\pm$ 0.22 & 0.39 $\pm$ 0.07 & 2.04 $\pm$ 0.00 & 
 Ly$\alpha$ & 1.080 \\
2535.87 $\pm$ 0.10 & 0.91 $\pm$ 0.06 & 2.04 $\pm$ 0.00 & 
 Ly$\alpha$ & 1.086 \\
2571.31 $\pm$ 0.36 & 0.52 $\pm$ 0.11 & 3.41 $\pm$ 0.86 & 
 Ly$\alpha$ & 1.115 \\
2586.93 $\pm$ 0.07 & 1.21 $\pm$ 0.08 & 2.16 $\pm$ 0.17 & 
 FeII 2586.65 (MW) &     32.39 \\
2594.62 $\pm$ 0.22 & 0.38 $\pm$ 0.07 & 2.04 $\pm$ 0.00 & 
 MnII 2594.50 (MW) &     14.27 \\
2600.20 $\pm$ 0.13 & 0.96 $\pm$ 0.10 & 2.73 $\pm$ 0.32 & 
 FeII 2600.17 (MW) &      2.84 \\
2616.63 $\pm$ 0.11 & 1.33 $\pm$ 0.10 & 3.17 $\pm$ 0.27 & 
 Ly$\alpha$ & 1.152 \\
2639.34 $\pm$ 0.09 & 1.00 $\pm$ 0.09 & 2.26 $\pm$ 0.23 & 
 Ly$\alpha$ & 1.171 \\
2667.22 $\pm$ 0.32 & 0.26 $\pm$ 0.07 & 2.04 $\pm$ 0.00 & 
 Ly$\alpha$ & 1.194 \\
2678.21 $\pm$ 0.39 & 1.08 $\pm$ 0.15 & 5.82 $\pm$ 0.93 & 
 Ly$\alpha$ & 1.203 \\
2695.68 $\pm$ 0.23 & 0.38 $\pm$ 0.07 & 2.02 $\pm$ 0.00 & 
 Ly$\alpha$ & 1.217 \\
\hline
\end{tabular}
\newpage
\begin{tabular}{|c|c|c|l|r|}
\multicolumn{5}{c}{\sc Table 2 (cont.)} \\
\hline \hline
$\lambda_{\rm obs}$ (\AA) & $W_{\lambda}$ (\AA) & FWHM (\AA) & 
\multicolumn{1}{c|}{ID} & $v$ or $z$ \\
\hline \hline
\multicolumn{5}{|l|}{QSO 1258$+$285 (cont.)} \\
\hline
2700.10 $\pm$ 0.07 & 1.26 $\pm$ 0.08 & 2.25 $\pm$ 0.17 & 
 Ly$\alpha$ & 1.221 \\
2706.63 $\pm$ 0.07 & 2.86 $\pm$ 0.10 & 4.25 $\pm$ 0.19 & 
 Ly$\alpha$ & 1.226 \\
2748.49 $\pm$ 0.22 & 1.37 $\pm$ 0.13 & 2.50 $\pm$ 0.54 & 
 Ly$\alpha$ & 1.261 \\
2768.53 $\pm$ 0.21 & 0.40 $\pm$ 0.06 & 2.04 $\pm$ 0.00 & 
 Ly$\alpha$ & 1.277 \\
2796.40 $\pm$ 0.12 & 0.69 $\pm$ 0.06 & 2.04 $\pm$ 0.00 & 
 MgII 2796.35 (MW) &      5.31 \\
2802.91 $\pm$ 0.09 & 1.91 $\pm$ 0.10 & 3.50 $\pm$ 0.21 & 
 MgII 2803.53 (MW) &    -66.79 \\
2810.36 $\pm$ 0.08 & 2.62 $\pm$ 0.10 & 4.34 $\pm$ 0.20 & 
 Ly$\alpha$ & 1.311 \\
2821.57 $\pm$ 0.37 & 0.18 $\pm$ 0.06 & 2.04 $\pm$ 0.00 & 
 Ly$\alpha$ & 1.321 \\
2843.77 $\pm$ 0.60 & 0.57 $\pm$ 0.11 & 6.47 $\pm$ 1.45 & 
 Ly$\alpha$ & 1.339 \\
2854.02 $\pm$ 0.10 & 0.89 $\pm$ 0.06 & 2.85 $\pm$ 0.23 & 
 MgI 2852.96 (MW) &    111.40 \\
2860.39 $\pm$ 0.09 & 1.85 $\pm$ 0.11 & 3.42 $\pm$ 0.21 & 
 Ly$\alpha$ & 1.352 \\
2864.76 $\pm$ 0.17 & 1.00 $\pm$ 0.11 & 3.38 $\pm$ 0.41 & 
 Ly$\alpha$ & 1.356 \\
2874.02 $\pm$ 0.15 & 0.66 $\pm$ 0.07 & 3.07 $\pm$ 0.35 & 
 Ly$\alpha$ & 1.364 \\
3076.39 $\pm$ 0.22 & 0.43 $\pm$ 0.09 & 2.14 $\pm$ 0.51 & 
 ... & ... \\
\hline
\multicolumn{5}{|l|}{QSO 1348$+$263} \\
\hline
2344.48 $\pm$ 0.15 & 0.62 $\pm$ 0.09 & 2.19 $\pm$ 0.37 & 
 FeII 2344.21 (MW) &     33.94 \\
2375.97 $\pm$ 0.76 & 0.82 $\pm$ 0.18 & 7.45 $\pm$ 1.94 & 
 FeII 2374.46 (MW) &    191.02 \\
2382.71 $\pm$ 0.09 & 0.83 $\pm$ 0.07 & 2.02 $\pm$ 0.00 & 
 FeII 2382.76 (MW) &     -6.67 \\
2418.64 $\pm$ 0.39 & 0.20 $\pm$ 0.06 & 2.04 $\pm$ 0.00 & 
 ... & ... \\
2432.60 $\pm$ 0.56 & 0.39 $\pm$ 0.12 & 3.85 $\pm$ 1.32 & 
 ... & ... \\
2543.10 $\pm$ 0.34 & 0.23 $\pm$ 0.06 & 2.04 $\pm$ 0.00 & 
 ... & ... \\
2586.62 $\pm$ 0.16 & 0.93 $\pm$ 0.09 & 3.35 $\pm$ 0.38 & 
 FeII 2586.65 (MW) &     -3.20 \\
2594.99 $\pm$ 0.37 & 0.28 $\pm$ 0.08 & 2.65 $\pm$ 0.88 & 
 MnII 2594.50 (MW) &     56.73 \\
2600.16 $\pm$ 0.09 & 0.75 $\pm$ 0.05 & 2.02 $\pm$ 0.00 & 
 FeII 2600.17 (MW) &     -1.91 \\
2632.52 $\pm$ 0.59 & 0.41 $\pm$ 0.11 & 4.60 $\pm$ 1.41 & 
 ... & ... \\
2745.79 $\pm$ 0.37 & 0.19 $\pm$ 0.06 & 2.04 $\pm$ 0.00 & 
 ... & ... \\
2796.45 $\pm$ 0.07 & 1.02 $\pm$ 0.05 & 2.04 $\pm$ 0.00 & 
 MgII 2796.35 (MW) &     10.93 \\
2803.41 $\pm$ 0.07 & 1.06 $\pm$ 0.07 & 2.27 $\pm$ 0.18 & 
 MgII 2803.53 (MW) &    -12.67 \\
3035.90 $\pm$ 0.63 & 0.37 $\pm$ 0.10 & 4.80 $\pm$ 1.50 & 
 ... & ... \\
3100.18 $\pm$ 0.33 & 0.25 $\pm$ 0.06 & 2.04 $\pm$ 0.00 & 
 ... & ... \\
\hline
\end{tabular}

\begin{tabular}{|l|c|c||c|c|c|c|c|c|}
\multicolumn{9}{c}{\sc Table 3} \\
\hline \hline
\multicolumn{3}{|c||}{} & \multicolumn{6}{c|}{log $N_{\rm lim}$ (cm$^{-2}$)} \\ \hline
Species & $\lambda$ (\AA) & $f$ & A 21 & A 151 & A 754 & A 1267 & A 1656 & A 1795 \\
\hline
AlI &  2264.16 & 0.133 & 13.08 & 13.08 & 13.13 & 12.69 & 13.19 & 13.10 \\
CaI &  2151.47 & 0.020 & 13.68 & 13.64 & 13.68 & 13.56 & 14.12 & 13.61 \\
FeII & 2382.77 & 0.343 & 12.64 & 12.62 & 12.53 & 12.41 & 12.94 & 12.51 \\
MgI &  2852.96 & 1.730 & 11.83 & 11.75 & 11.76 & 11.53 & 12.83 & 11.63 \\
MgII & 2796.35 & 0.629 & 12.28 & 12.18 & 12.21 & 11.99 & 12.51 & 12.11 \\
MnII & 2576.88 & 0.351 & 12.43 & 12.40 & 12.38 & 12.19 & 12.40 & 12.30 \\
NaI &  2853.72 & 0.002 & 14.96 & 14.90 & 14.89 & 14.68 & 14.89 & 14.83 \\
SiI &  2515.07 & 0.236 & 12.87 & 12.82 & 12.85 & 12.65 & 13.46 & 12.78 \\
\hline
\end{tabular}

\clearpage

\par\noindent
{\bf Figure 1.}  Calibrated HST/FOS spectra of six QSOs projected behind
clusters.  Axes are flux in units of $10^{15}$\ ergs/s/cm$^2$/\AA\ vs.\
wavelength in \AA.  The upper line traces the QSO ``continuum'' fit, while
the lower line shows the 3-sigma error spectrum which is propagated through
the HST reduction pipeline along with the science data.  QSO emission lines
are identified, and strong Galactic absorption lines are denoted with
dotted lines.  Tick marks denote special absorption features discussed in
the text.  Regions marked with a bold line are locations we expect to see
absorption from FeII or MgII in the cluster.

\vspace{1em}
\par\noindent
{\bf Table 1.}  Summary of the six observations and cluster/QSO
characteristics.  Five of the spectra were observed for this program;
data for Coma were obtained from the HST archive.  ${\dot M}$ is the
cooling rate; it is zero for non-cooling flow clusters.

\vspace{1em}
\par\noindent
{\bf Table 2.}  All detected lines, showing line center, equivalent width,
probable ID, and velocity in \kps\ (for Galactic components, identified as
``MW'') or redshift (if of extragalactic origin).  Uncertainties reported
are 1-$\sigma$ errors.

\vspace{1em}
\par\noindent
{\bf Table 3.}  Upper limits to the column densities of various species
along each cluster line of sight.  Also listed is the strongest transition
for each species in the observed wavelength range, along with its
$f$-value.  The FeII and MgII transitions are the strongest lines expected
from low-ionization absorbing gas.

\end{document}